# Surface-enhanced Raman scattering in graphene deposited on Al$_x$Ga$_{1-x}$N/GaN axial heterostructure nanowires

Jakub Kierdaszuk[a,*], Mateusz Tokarczyk [a], Krzysztof M. Czajkowski [a], Rafał Bożek [a],

Aleksandra Krajewska [b,c], Aleksandra Przewłoka [b], Wawrzyniec Kaszub [b],

Marta Sobanska [d], Zbigniew R. Zytkiewicz [d], Grzegorz Kowalski [a],

Tomasz J. Antosiewicz [a], Maria Kamińska [a], Andrzej Wysmołek [a], Aneta Drabińska [a]

[a] *Faculty of Physics, University of Warsaw, Pasteura 5, Warsaw, Poland*

[b] *Institute of Electronic Materials Technology, Wólczyńska 133, Warsaw, Poland*

[c] *Institute of Optoelectronics, Military University of Technology, Kaliskiego 2, Warsaw, Poland*

[d] *Institute of Physics, Polish Academy of Sciences, Lotników 32/46, Warsaw, Poland*

*Corresponding author. Tel. +48 573931438, E-mail: jakub.kierdaszuk@fuw.edu.pl

## ABSTRACT

The surface-enhanced Raman scattering in graphene deposited on Al$_x$Ga$_{1-x}$N/GaN axial heterostructure nanowires was investigated. The intensity of graphene Raman spectra was found not to be correlated with aluminium content. Analysis of graphene Raman bands parameters, KPFM and electroreflectance showed a screening of polarization charges. Theoretical calculations showed that plasmon resonance in graphene is far beyond the Raman spectral range. This excludes the presence of an electromagnetic mechanism of SERS and therefore suggests the chemical mechanism of enhancement.

## INTRODUCTION

Studies of graphene deposited on vertically aligned gallium nitride nanowires (GaN NWs) are interesting not only because of possible applications in photovoltaics but also because of the presence of surface-enhanced Raman scattering in graphene.[1],[2],[3] Intensity of Raman bands has been found to be more than one order of magnitude higher than in graphene deposited on GaN epilayer. Our previous results have excluded light interference in NWs layer to be responsible for this enhancement.[4] However, they have not explained which of two mechanisms of enhancement: electromagnetic or chemical dominates. The electromagnetic mechanism is related to the formation of surface plasmons in graphene due to the existence of polarization charges on top of nanowires.[5] The chemical mechanism assumes that Raman spectra enhancement is caused by the charge-transfer resulting from chemical bonding at the GaN/graphene interface. AlGaN is a wide bandgap semiconductor of the wurtzite structure. Due to the spontaneous and piezoelectric polarization along the c-axis, a high concentration of polarization charges is present on its surface.[6],[7] Polarization charges are usually screened by free carriers. However, recent works showed that NW structure reduces free carrier screening effect which for example enable to create piezo-generator integrating a vertical array of GaN NWs.[8] Total electric polarization in AlGaN consists of spontaneous polarization ($P_{SP}$) and of piezoelectric polarization ($P_{PE}$). Both polarization components increase with aluminium content ($x$) in $Al_xGa_{1-x}N$.[6] However, the detailed discussion of actual polarization charge density on NWs top surface needs to be extended by calculating the exact value of piezoelectric polarization using the lattice parameters which can be obtained from X-ray diffraction. An additional technique that can give information about the electric field present in the structure is electroreflectance spectroscopy.[9],[10] Observation of Franz-Keldysh oscillations in electroreflectance spectra provides direct evidence of the presence of electric fields caused by polarization charges.[11] On the other hand energies and widths of graphene G and 2D bands depend both on carrier concentration and also on graphene strain.[12],[13] So, Raman micromapping allows to trace how AlGaN NWs locally gate graphene and change its carrier concentration and strain.[4] Furthermore, Kelvin Probe Force Microscopy (KPFM) were applied in order to measure the topography of graphene deposited on $Al_xGa_{1-x}N$ NWs and distribution of potential on the graphene surface. Value of potential in graphene is proportional to the value carrier concentration.[14] Therefore KPFM measurements enable to investigate how different NW surface effect on carrier concentration in graphene and consequently their nanogating in better resolution than Raman spectroscopy.[4] Thus, studies of graphene deposited on $Al_xGa_{1-x}N$ NWs with different composition should give an insight on how electric field induced in NWs influences SERS effect in graphene, and furthermore should recognize which enhancement mechanism is dominant.

EXPERIMENTAL DETAILS

GaN NWs were grown by Plasma Assisted Molecular Beam Epitaxy on Si(111) substrate under N-rich conditions without use of any catalyst.[15] After a growth of 900 nm of GaN NWs, Al source was opened and 100 nm of $Al_xGa_{1-x}N$ was grown. The crystallographic orientation of NWs was (000-1). During our measurements, several samples with *x* content varying from 0 to 1were studied (named from Al0 to Al100 respectively). Graphene was grown by Chemical Vapour Deposition on Cu foil by methane precursor.[16] Polymer frame was used to transfer graphene onto NW substrates.[17] Scanning electron microscopy (SEM) measurements were performed on a microscope integrated with Helios Nanolab 600 Focus Ion Beam. It was equipped with a secondary electron detector at 5 kV electron beam voltage. Phillips X-pert diffractometer equipped with a Cu sealed tube X-ray source, a four bounce Ge (220) Bartels monochromator and a channel-cut Ge (220) analyzer was used in High-resolution X-ray diffraction (HRXRD) measurements. Electroreflectance spectroscopy relays on the measurement of reflectivity during external modulation of the electric field in the structure. In our case the AC voltage was applied between the graphene and Si substrate, assuring the modulation of the electric field in NWs. The sample was illuminated by the monochromatic light and the reflected light was focused onto a silicon diode. The DC and AC signal was measured by the voltmeter and lock-in amplifier respectively. As a light source, 75 W Xe lamp was used. Raman measurements were performed by T64000 Horiba Jobin-Yvon spectrometer with a 532 nm excitation Nd:YAG laser. Raman micromapping for each sample was performed on an area of a 9 μm², with 100 nm step using 100× microscope objective with a spatial resolution of approximately 300 nm. KPFM measurements were performed using Digital Instruments Multimode Atomic Force Microscope with a needle of 50 nm diameter, capable of measuring the electric potential of the specimen. In KPFM measurements, due to the inability to determine the reference potential, values of the measured potential are arbitrary, however, their values and local variations can be compared between the investigated samples.

EXPERIMENTAL RESULTS

SEM measurements show a low roughness of graphene on NWs (the representative image was presented in Fig. 1). Small wrinkles are caused by an expansion of graphene hanging between nearest NWs. SEM also confirmed the similar value of NWs height and density distribution on the surface in all investigated samples. HRXRD measurements show peaks related to symmetric (002) and asymmetric (105) GaN and $Al_xGa_{1-x}N$ reflections (Fig. 2a, b). Due to the strain, peaks are shifted from the literature values of unstrained GaN and AlGaN.[6] Values of calculated relative strain are presented in Table 1. Spontaneous polarization in AlGaN can be calculated using the following formula:[6]

$$P_{SP} = -0.052x - 0.029. \qquad (2)$$

where $x$ is aluminium content while piezoelectric polarization $P_{PE}$ depends both on $x$ and strain: [6]

$$P_{PE}(x) = (0.73 + 0.73x)\epsilon_{a\%} + 2(-0.11x - 0.49)\epsilon_{c\%}. \qquad (1)$$

The calculated value of $P_{PE}$ in the case of our samples is much smaller than $P_{SP}$ (Table 1), and no explicit correlation between $P_{PE}$ and $x$ is observed. Total polarization is a sum of spontaneous and piezoelectric contributions. Non-zero polarization in AlGaN NWs results in presence of positive polarization charge on top of the NWs with sheet polarization charge concentration ($n_s$) equal to:

$$n_s = |P_{SP} + P_{PE}|/e. \qquad (3)$$

In case of our samples, $n_s$ is increasing with $x$ and varies from $1.2 \cdot 10^{13}$ cm$^{-2}$ for Al0 sample to $4.1 \cdot 10^{13}$ cm$^{-2}$ for Al100 sample (Table 1).

The direct verification of the polarization charge presence can be performed by electroreflectance measurements presented in Fig. 3. Due to the spectral range, this technique was only available for samples with the lowest Al content (below 20%). The electroreflectance signal shape depends on the strength of the electric field present in the structure. In the low-field limit, when the energy of the accelerated particle is smaller than energy broadening, third-derivative line shapes appear. In practice, for GaN heterostructures at room temperatures, this limit is realized if the electric field is lower than 20 kV/cm. If the energy of the accelerated particle is larger than broadening energy it is possible to observe very characteristic Franz-Keldysh oscillations for energies larger than energy band gap of the material. The period of these oscillations is proportional to the electric field and the amplitude vanishes exponentially and depends not only on the electric field but also on the broadening energy.[11] In each sample, two signals were observed (Fig. 3). One, at a wavelength of about 360 nm, corresponding to the energy gap of GaN. The second line corresponded to the reflectivity from AlGaN cap. It changed its position from about 355 nm for 3% of aluminium content to about 330 nm for 15% of aluminium content. Both signals have shape characteristic for the low electric field limit and no sign of Franz-Keldysh oscillations is observed. Therefore, we can conclude that for low aluminium content the polarization charges are screened by the free carriers and no electric field in AlGaN caps is present. This result casts doubt on whether NW structure indeed reduces polarization charge screening as was reported in the literature. [8] To trace the real influence of NWs substrate on graphene Raman spectroscopy was performed. Representative Raman spectra of graphene on NWs are presented in Fig. 4a. Intensities of graphene bands in each sample are enhanced in contrast to the reference sample of graphene on GaN epilayer. However, analysis of the enhancement factors for each band shows no explicit correlation with aluminium content (Fig. 4b). To trace how NW substrate affects graphene strain and carrier concentration statistical analysis of G and 2D band parameters: energies ($E_G$, $E_{2D}$), Full Width at Half Maximum ($F_G$, $F_{2D}$) and their intensity ratio ($R_{2DG}$) was performed. [13],[12] No explicit correlation for any of this parameter with $x$ is observed (Fig. 5a,b,c,d,e). This suggests that graphene deposited on NWs with similar density distribution is strained in a similar way and aluminium content in AlGaN caps does not significantly impact average carrier concentration in graphene. In order to final clarification how different NW substrate effect value of carrier concentration with nanometre resolution, Kelvin Probe Atomic Force Microscopy was applied. Local graphene self-induced nanogating by NW substrate was previously reported.[4] Topography images confirmed low roughness of the graphene surface observed in SEM images (Fig. 6a-f). Spatial modulation of potential in whole investigated samples was observed on the measured surface (Fig. 6f-j). Analysis of potential profiles enables us to estimate the value of potential modulation in graphene on axial heterostructure NWs. The average value of nanowire-induced potential modulation in graphene varies from 0.8 mV for Al0 sample to 3.7 mV for Al100 sample. For samples Al25-Al75 that parameter is equal to 3.4 mV, 3.4 mV and 2.8 mV respectively. Therefore, no explicit correlation between aluminium content and the value of potential modulation is observed. Thus, KPFM results confirm Raman spectroscopy results that aluminium content generally does not affect the carrier concentration in graphene.

Experimental results show a screening of polarization charges in NW structures and thus they question the electromagnetic mechanism of enhancement and turn to the chemical one. In order to confirm this, numerical simulations were applied. Calculations performed by finite difference time domain method show the presence of electric field enhancement in the structure of randomly distributed NWs (Fig. 7a). However, the value of enhancement is lower than observed in typical metallic plasmonic structures. What's more, our calculations show the absence of plasmonic absorption in the investigated spectral range. They suggest that calculated enhancement can be caused by light interference (Fig. 7b). However, the interference effect was excluded in our previous results since the laser line is in the minimum of interference and the intensity ratios of individual Raman band do not follow the interference spectrum (Fig. 4b).[4] This discrepancy can be understood by taking into account that simulation was performed on randomly distributed NWs of equal height while on SEM images it can be seen that the real roughness of NWs substrate is low but still visible. Therefore, experimental results and numerical simulations show that electromagnetic mechanism is not responsible for SERS effect in graphene on NWs and suggest the occurrence of a chemical mechanism.

CONCLUSIONS

We observed the surface-enhanced Raman scattering effect in graphene deposited on AlGaN NWs with different aluminium content. We calculated the value of total polarization in AlGaN and found out that the value of sheet polarization charge concentration on top of NWs is correlated with Al content. Interestingly, different amount of Al does not affect enhancement factors of graphene Raman bands. The detailed analysis of Raman bands showed the presence of screening of polarization charges. This was confirmed by KPFM and electroreflectance results. Therefore, experimental results question the possibility of the electromagnetic mechanism being responsible for the observed SERS. Additionally, numerical calculations also excluded the electromagnetic mechanism of the observed enhancement of graphene Raman spectra. Therefore, our results suggest a chemical mechanism to be responsible for the observed enhancement.

ACKNOWLEDGEMENTS

This work was partially supported by the Ministry of Science and Higher Education in years

2015-2019 as a research grant "Diamond Grant" (No. DI2014 015744)." Financial support from the Polish Nacional Science Centre under grants 2016/21/N/ST3/03381 and 2016/23/B/ST7/03745 as well as by the European Union within European Regional Development Fund, grant POIG.01.03.01-00-159/08 (InTechFun) is acknowledged.

LIST OF FIGURES

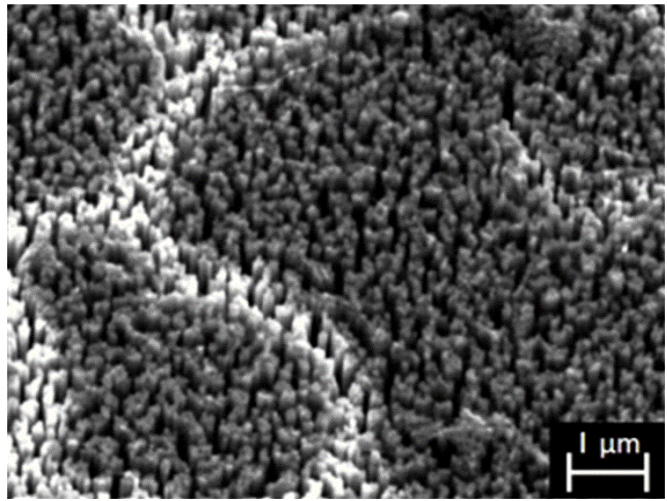

**Figure 1.** Representative SEM image of graphene deposited on GaN NWs with $Al_{0.75}Ga_{0.25}N$ caps.

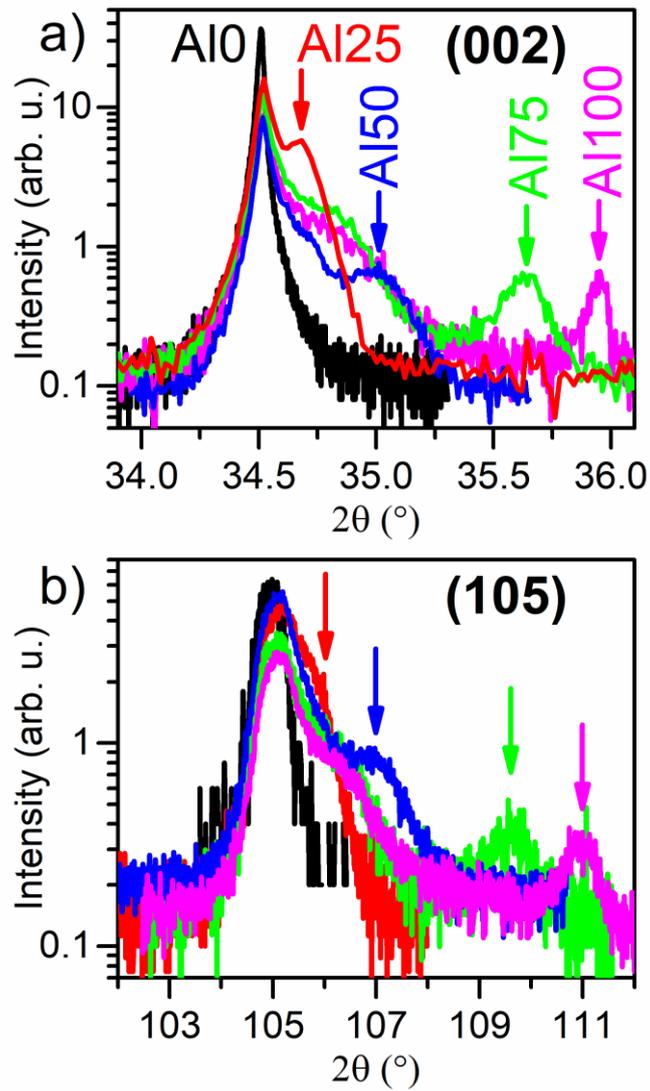

**Figure 2.** HRXRD intensity recorded around: a) (002) reflection, b) (105) reflection

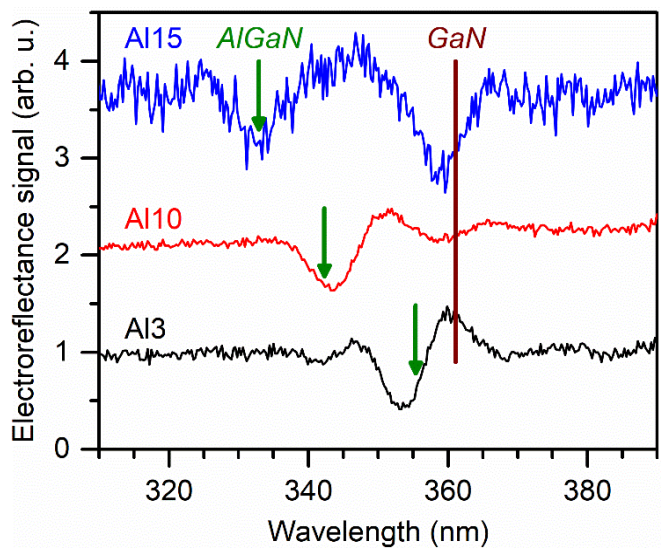

**Figure 3.** Electroreflectance spectra of investigated samples

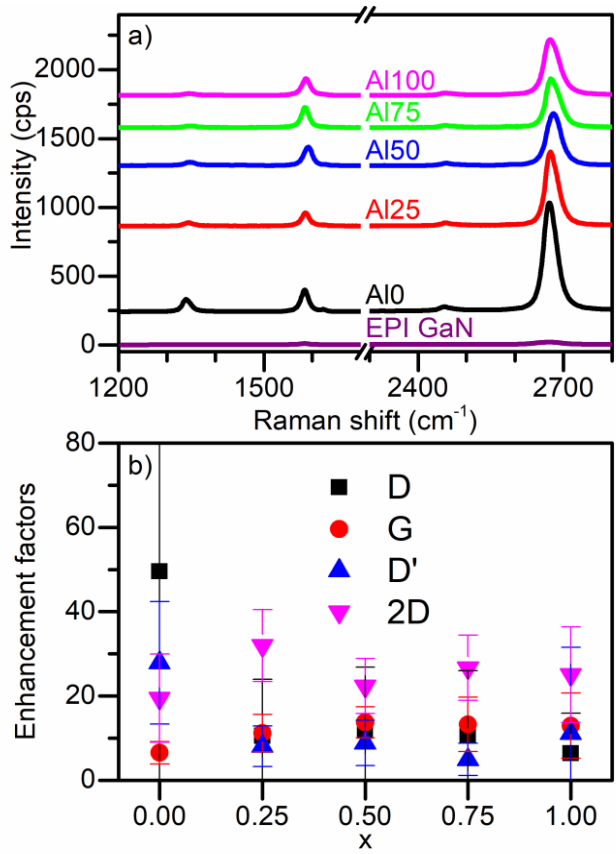

**Figure 4.** a) Raman spectra of investigated samples and reference samples of graphene on GaN epilayer, b) enhancement factors for each Raman bands.

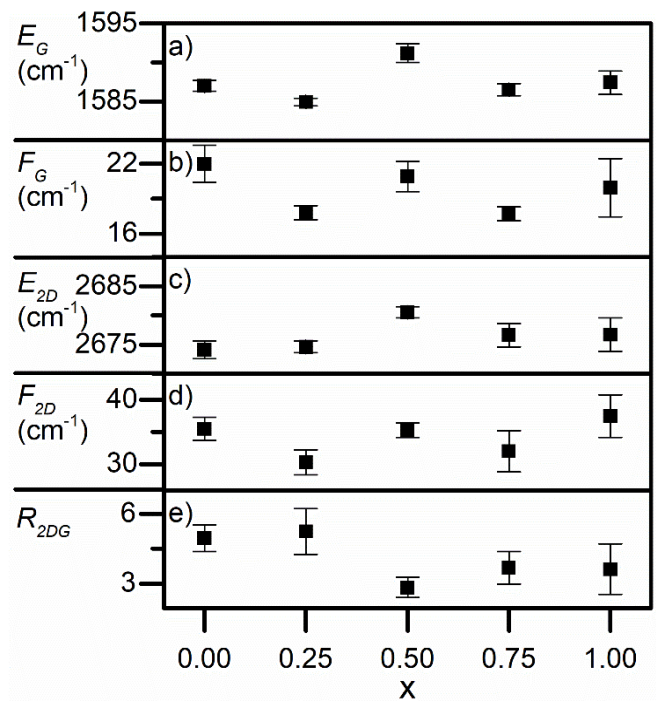

**Figure 5.** Average values of: a) $E_G$, b) $F_G$, c) $E_{2D}$, d) $F_{2D}$, e) $R_{2DG}$ for different NWs composition.

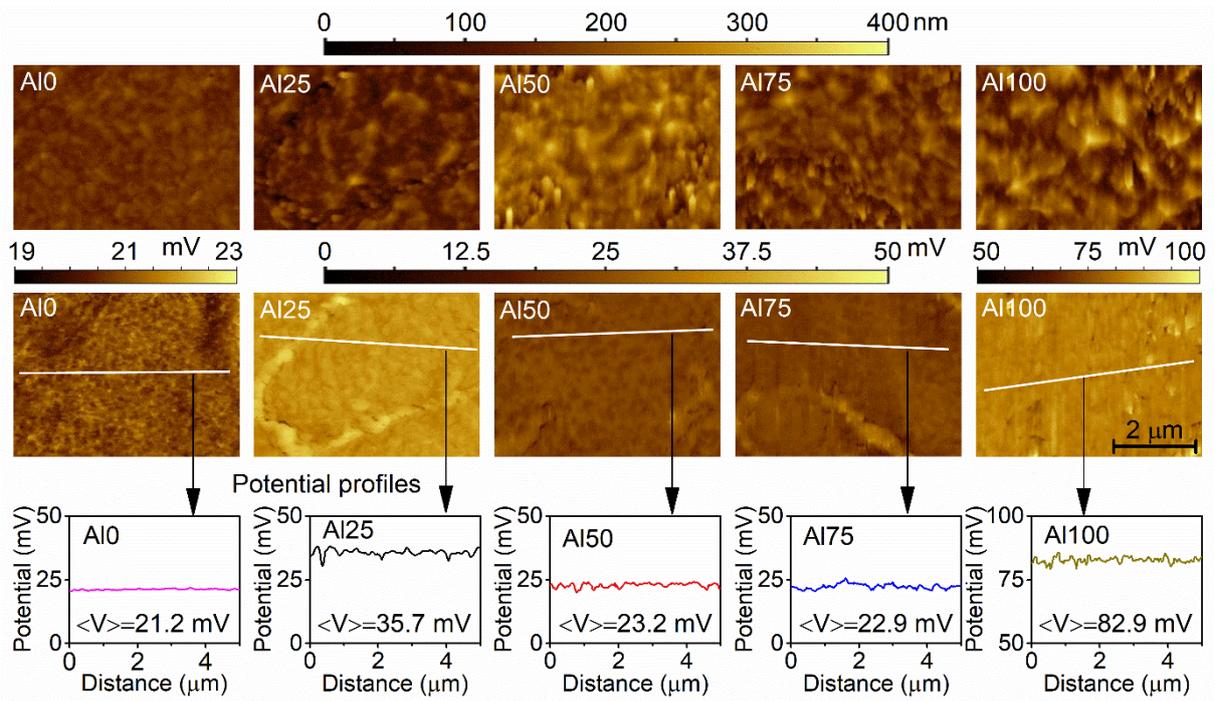

**Figure 6.** KPFM images of topography in investigated samples (a-e), potential distribution (f-j) and representative potential profiles (k-o).

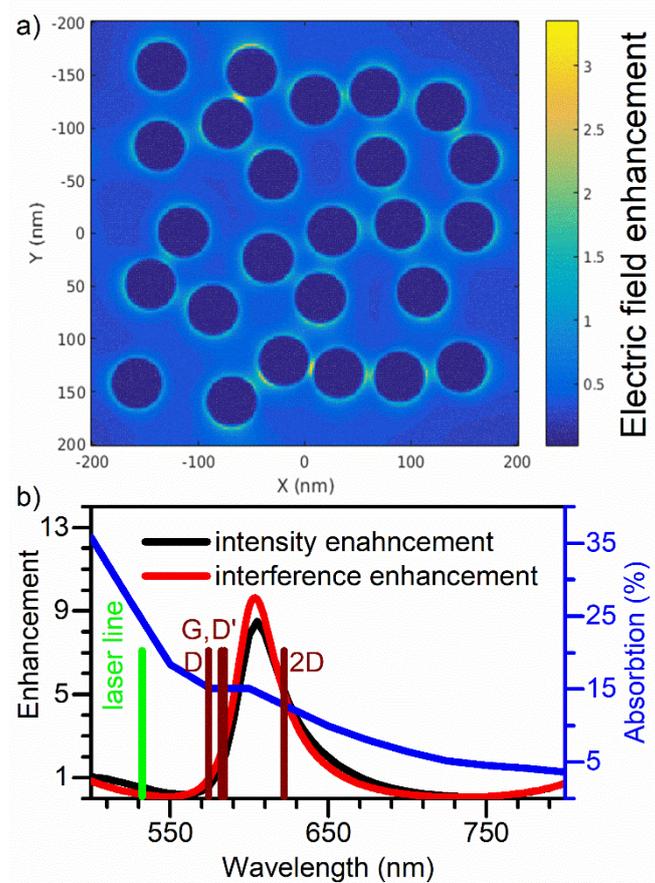

**Figure 7.** a) 2D map of electric field enhancement in graphene deposited on GaN NWs, b) dependence of electric field enhancement, absorption and interference intensity enhancement on wavelength.

LIST OF TABLES

**Table 1.** Values of lattice constants (*a, c*) for GaN and AlGaN and strain in the top layer ($\varepsilon_a$, $\varepsilon_c$) obtained from HRXRD measurements, calculated values of piezoelectric and spontaneous polarization ($P_{PE}$, $P_{SP}$) and concentration of polarization charge ($n_s$) present on Al$_x$Ga$_{1-x}$N NWs top surface.

| x | 0 | 0.25 | 0.5 | 0.75 | 1 |
|---|---|---|---|---|---|
| $c_{GaN}$ (Å) | 5.194 | 5.193 | 5.193 | 5.192 | 5.193 |
| $a_{GaN}$ (Å) | 3.163 | 3.157 | 3.141 | 3.143 | 3.137 |
| $c_{AlGaN}$ (Å) | - | 5.170 | 5.132 | 5.029 | 4.992 |
| $a_{AlGaN}$ (Å) | - | 3.148 | 3.124 | 3.109 | 3.078 |
| $\varepsilon_a$ (%) | 0.17 | 0.70 | 0.96 | -0.08 | 0.20 |
| $\varepsilon_c$ (%) | -0.80 | -0.69 | -0.85 | -0.71 | -1.10 |
| $P_{PE}$ (Cm$^{-2}$) | 0.009 | 0.013 | 0.020 | 0.007 | 0.016 |
| $P_{SP}$ (Cm$^{-2}$) | -0.029 | -0.042 | -0.055 | -0.068 | -0.081 |
| $n_s$ (cm$^{-2}$) | 1.2·10$^{13}$ | 1.8·10$^{13}$ | 2.2·10$^{13}$ | 3.8·10$^{13}$ | 4.1·10$^{13}$ |